# $Na_{1-x}Sn_2P_2$ as a new member of van der Waals-type layered tin pnictide superconductors

Yosuke Goto,[1,*] Akira Miura,[2] Chikako Moriyoshi,[3] Yoshihiro Kuroiwa,[3] Tatsuma D. Matsuda,[1] Yuji Aoki,[1] and Yoshikazu Mizuguchi[1]

1) Department of Physics, Tokyo Metropolitan University, 1-1 Minami-osawa, Hachioji, Tokyo 192-0397, Japan
2) Faculty of Engineering, Hokkaido University, Kita 13, Nishi 8 Sapporo 060-8628, Japan
3) Department of Physical Science, Hiroshima University, 1-3-1 Kagamiyama, Higashihiroshima, Hiroshima 739-8526, Japan

*E-mail: y_goto@tmu.ac.jp

**Abstract**

Superconductors with a van der Waals (vdW) structure have attracted a considerable interest because of the possibility for truly two-dimensional (2D) superconducting systems. We recently reported $NaSn_2As_2$ as a novel vdW-type superconductor with transition temperature ($T_c$) of 1.3 K. Herein, we present the crystal structure and superconductivity of new material $Na_{1-x}Sn_2P_2$ with $T_c$ = 2.0 K. Its crystal structure consists of two layers of a buckled honeycomb network of SnP, bound by the vdW forces and separated by Na ions, as similar to that of $NaSn_2As_2$. Amount of Na deficiency ($x$) was estimated to be 0.074(18) using synchrotron X-ray diffraction. Bulk nature of superconductivity was confirmed by the measurements of electrical resistivity, magnetic susceptibility, and specific heat. First-principles calculation using density functional theory shows that $Na_{1-x}Sn_2P_2$ and $NaSn_2As_2$ have comparable electronic structure, suggesting higher $T_c$ of $Na_{1-x}Sn_2P_2$ is resulted from increased density of states at the Fermi level due to Na deficiency. Because there are various structural analogues with tin-pnictide (SnPn) conducting layers, our results indicate that SnPn-based layered compounds can be categorized into a novel family of vdW-type superconductors, providing a new platform for studies on physics and chemistry of low-dimensional superconductors.

**Introduction**

Superconducting behavior with exotic characteristics is often observed in materials with a layered two-dimensional crystal structure. Low dimensionality affects the electronic structure of these materials, potentially leading to high transition temperatures ($T_c$) and/or unconventional pairing mechanisms (1,2). Among the layered superconductors, much attention has been paid to the van der Waals (vdW) materials because of the possibility for truly two-dimensional (2D) superconducting systems (3–5). Owing to the recent development on the mechanical exfoliation techniques, various vdW materials are found to be suitable to make a 2D system by reducing their thickness down to the level of individual atomic layers (6). As an example, atomically-thin $NbSe_2$ crystals turn out to host unusual superconducting states, including Ising superconductivity with a strong in-plane upper critical field (4) and a field-induced Bose-metal phase under the out-of-plane magnetic field (5). In order to clarify the underlying mechanisms of such exotic states and to investigate whether or not they are generic, further studies, particularly on different types of vdW superconductors, are highly desirable.

We recently reported the discovery of $NaSn_2As_2$ superconductor with $T_c$ = 1.3 K (7). $NaSn_2As_2$ crystallizes in a trigonal $R\bar{3}m$ unit cell, consisting of two layers of a buckled honeycomb network of SnAs, bound by the vdW forces and separated by Na ions (8), as schematically shown in Figure 1a and 1c. Because of the vdW gap between the SnAs layers, it can be readily exfoliated through both mechanical and liquid-phase methods (8,9). Besides, the sister compound $SrSn_2As_2$, having a crystal structure analogous to $NaSn_2As_2$, has been theoretically suggested to be very close to the topological critical point, hosting three-dimensional Dirac state at the Fermi level (10), which was experimentally investigated by angle-resolved photoemission spectroscopy (11). There are various structural analogues with conducting tin-pnictide (SnPn) layers, including $Sn_4Pn_3$ (12,13) and ASnPn (14–18), as well as $ASn_2Pn_2$ (8,9,11,19–21), where A denotes alkali or alkaline earth metal (see Fig. 1). Indeed, $Sn_4Pn_3$ was reported to be a superconductor with $T_c$ = 1.2–1.3 K (22,23), although detailed superconducting characteristics have not been reported. In addition to these superconductors, ASnPn is attractive for thermoelectric application because of its relatively low lattice thermal conductivity lower than 2 $Wm^{-1}K^{-1}$ at 300 K, most likely due to lone-pair effects (15). These results strongly suggest that SnPn-based layered compounds can be regarded as a novel family of vdW-type compounds exhibiting various functionality.

Herein, we report $Na_{1-x}Sn_2P_2$ as a new member of SnPn-based vdW-type superconductors with $T_c$ = 2.0 K. Crystal structure analysis was performed using synchrotron powder X-ray diffraction (SPXRD). Superconducting properties were examined by the measurements of the electrical resistivity ($\rho$), magnetic susceptibility ($\chi$) and the specific heat ($C$). Electronic structure was calculated on the basis of density functional theory (DFT).

**Results and Discussion**
**Crystal structure analysis**

Figure 2 shows the SPXRD pattern and the Rietveld fitting results for $Na_{1-x}Sn_2P_2$. Almost all the diffraction peaks can be assigned to those of the trigonal $R\bar{3}m$ (No. 166) space group, indicating that $Na_{1-x}Sn_2P_2$ is isostructural to $NaSn_2As_2$. Although diffraction peaks attributable to elemental Na (10.1 wt%) was also observed, Na does not show superconductivity at least under ambient pressure. The results of the Rietveld analysis including the refined structural parameters were listed in Table 1. The lattice parameters were $a$ = 3.8847(2) Å and $c$ = 27.1766(13) Å. These are smaller than those of $NaSn_2As_2$ ($a$ = 4.00409(10) Å and $c$ = 27.5944(5) Å), mainly because of smaller ionic radius of P ions than As ions. The site occupancy of Na site was evaluated to be 0.926(18), suggesting that the sample in the present study contains Na deficiency. Note that energy dispersive X-ray spectroscopy is not suitable to evaluate the chemical composition of the present sample because elemental Na is also observed as impurity phase.

**Superconducting properties**

Figures 3a and 3b show the $\rho$–$T$ plots for polycrystalline $Na_{1-x}Sn_2P_2$. Metallic behavior of the electrical resistivity was observed at temperatures above 10 K. A sharp drop in $\rho$ was observed at 2.0 K, accompanied by zero resistivity at temperatures under 1.9 K, which indicates a transition to superconducting states. The transition temperature shifted toward lower temperatures with increasing applied magnetic field, as shown in Fig. 3c. It is noteworthy that the superconducting transition was distinctly broadened under magnetic field, probably because of the anisotropic upper critical field due to the two-dimensional layered crystal structure. The transition temperatures, $T_c^{90\%}$ and $T_c^{zero}$, obtained from the temperature dependences of electrical resistivity under magnetic fields are shown in Fig. 3d. Here, $T_c^{90\%}$ is defined as the temperature at which $\rho$ is at 90% of the value at 3 K (normal state resistivity just above $T_c$), as indicated by a dashed line in Fig. 3c. The dependence of the upper critical field ($H_{c2}$) on temperature is still almost linear at $T \approx 0.5$ K. Namely, the curve deviates from the Werthamer–Helfand–Hohemberg (WHH) model (24). Here, the Pauli paramagnetic effect should be negligible because the Pauli limiting field is estimated as $1.84 \times T_c = 3.7$ T. We estimate $\mu_0 H_{c2}(0)$ as 1.5–1.6 T using linear extrapolation of $H_{c2}$– $T_c^{90\%}$ plot. The coherence length $\xi$ was estimated to be ~15 nm using the equation of $\xi^2 = \Phi_0/2\pi\mu_0 H_{c2}$, where $\Phi_0$ is magnetic flux quantum.

Figure 4 shows $T$ dependence of magnetization ($M$) for Na$_{1-x}$Sn$_2$P$_2$. Diamagnetic signals corresponding to superconducting transition was observed below 1.9 K, consistent with zero resistivity in $\rho$–$T$ data. It should be noted that weak diamagnetic signal is also seen at around 3.7 K, probably due to trace Sn, although resistivity and specific heat (see below) does not show any anomaly at this temperature

Figure 5a shows $C/T$ as a function of $T^2$. A steep jump in $C/T$ is observed at around 1.7 K, which is in reasonable agreement with the superconducting transition observed in the resistivity and magnetization. Because observed lattice specific heat for Na$_{1-x}$Sn$_2$P$_2$ in the normal state deviates from simple Debye model, the experimental data were fitted with a function including Einstein model:

$$C = \gamma T + \beta T^3 + C_{Einstein}$$

$$C_{Einstein} = A \cdot 3N_A k_B \left(\frac{\Theta_E}{T}\right)^2 \exp\left(\frac{\Theta_E}{T}\right)\left(\exp\left(\frac{\Theta_E}{T}\right) - 1\right)^{-2}$$

where $\gamma$ is the Sommerfeld coefficient, $\beta$ is a phonon specific heat parameter, $\Theta_E$ is a characteristic temperature of the low-energy Einstein mode, $N_A$ is the Avogadro constant, $k_B$ is the Boltzmann constant, and A is fitting parameter. The fit yields $\gamma = 5.31$ mJmol$^{-1}$K$^{-2}$, $\beta = 0.73$ mJmol$^{-1}$K$^{-4}$, A = 0.0095, and $\Theta_E = 34$ K. Considering the number of Einstein mode is $3AN_A$, the number of the acoustical mode is $3(n-A)N_A$, where n is the number of atoms per formula unit. Accordingly, the Debye temperature ($\Theta_D$) is represented as $(12\pi^4(n-A)N_A k_B/5\beta)^{1/3}$. We evaluated $\Theta_D$ of Na$_{1-x}$Sn$_2$P$_2$ to be 237 K. As shown in Fig. 5b, the electronic specific heat jump at $T_c$ ($\Delta C_{el}$) is 9.15 mJmol$^{-1}$K$^{-2}$. From the obtained parameters, $\Delta C_{el}/\gamma T_c$ is calculated as 1.0, which is slightly lower but in reasonable agreement with the value expected from the weak-coupling BCS approximation ($\Delta C_{el}/\gamma T_c = 1.43$). The electron–phonon coupling constant ($\lambda$) can be determined by Macmillan's theory (25), which gives

$$\lambda = \frac{1.04 + \mu^* \ln(\Theta_D/1.45T_c)}{(1 - 0.62\mu^*)\ln(\Theta_D/1.45T_c) - 1.04}$$

where $\mu^*$ is defined as the Coulomb pseudopotential. Taking $\mu^* = 0.13$ gives $\lambda = 0.40$, which is consistent with weakly-coupled BCS superconductivity. Because the electron–phonon coupling constant of NaSn$_2$P$_2$ is comparable to that of NaSn$_2$As$_2$ ($\lambda = 0.44$), higher $T_c$ of NaSn$_2$P$_2$ with respect to NaSn$_2$As$_2$ is likely due to increased density of states at the Fermi energy and/or the Debye temperature. Indeed, the $\gamma$ and $\Theta_D$ of NaSn$_2$As$_2$ were evaluated to be 3.97 mJ mol$^{-1}$ K$^{-2}$ and 205 K, respectively (7). It should be noted that A = 0.0095 of Na$_{1-x}$Sn$_2$P$_2$ is distinctly lower than that of the compounds containing rattling atoms, such as *β*-pyrochlore *Ae*Os$_2$O$_6$ (*Ae* = Rb, Cs), where A = 0.34–0.47 (26). The deviation of lattice specific heat from simple Debye model in Na$_{1-x}$Sn$_2$P$_2$ suggests the existence of low-energy phonon excitations with the flat dispersion in a limited region of the reciprocal space, rather than rattling motion of atoms. Indeed, calculated phonon dispersion of isostructural compound NaSn$_2$As$_2$ shows nonlinear characteristics resulting from overlapping between acoustic and optical modes, most likely due to the existence of lone-pair electrons (15).

Figure 6 shows the calculated partial density of states of stoichiometric NaSn$_2$Pn$_2$ (Pn = P, As). Generally speaking, electronic structure of NaSn$_2$P$_2$ and NaSn$_2$As$_2$ is almost comparable. The energy bands from −12 eV to −10 eV and from −8 eV to −4 eV are mainly Pn s-orbitals and Sn s-orbitals in character, respectively. The bands that span from −4 eV to the Fermi energy are mainly Pn p-orbitals and Sn s/p-orbitals in character, confirming the electrical conduction is dominated by a SnPn covalent bonding network. The larger DOS of Pn p-orbitals than that of Sn p-orbitals in this energy region are consistent with the greater electronegativity of Pn. The energy bands mainly consisting of Sn s-orbitals are broadened, which is most likely due to the interlayer bonding. Na s-orbitals mainly locates from 1 eV to 3 eV, indicating the electron transfer from cationic Na layer to anionic SnPn layer. From the calculated electronic structure, it is evident that density of states at the Fermi energy is increased by Na deficiency, which reduces the Fermi energy. This is in agreement with higher $T_c$ of Na$_{1-x}$Sn$_2$P$_2$ with respect to NaSn$_2$As$_2$.

Very recently, studies on temperature-dependent magnetic penetration depth (34) and thermal conductivity (35) show that superconductivity of NaSn$_2$As$_2$ is fully gapped *s*-wave state in the dirty limit, which should be consistent with above mentioned scenario. Detailed investigation on effect of off-stoichiometry in these compounds is currently under investigation.

**Conclusion**

In summary, we present the crystal structure, electronic structure, and superconductivity of novel material Na$_{1-x}$Sn$_2$P$_2$. Structural refinement using SPXRD shows that crystal structure of Na$_{1-x}$Sn$_2$P$_2$ belongs to the trigonal $R\bar{3}m$ space group. Amount of *x* was estimated to be 0.074(18) from the Rietveld refinement. DFT calculations of the electronic structure confirm that the electrical conduction is dominated by a SnP covalent bonding network. Measurements of electrical resistivity, magnetic susceptibility, and specific heat confirm the bulk nature of superconductivity with $T_c$ = 2.0 K. On the basis of the structural and superconductivity characteristics of $Na_{1-x}Sn_2P_2$, which are similar to those of the structural analogue $NaSn_2As_2$, we consider that the SnPn layer can be a basic structure of layered superconductors. Because there are various structural analogues with SnPn-based conducting layers, our results indicate that SnPn-based layered compounds can be categorized into a novel family of vdW-type superconductors, providing a new platform for studies on physics and chemistry of low-dimensional superconductors.

**Methods**

Polycrystalline $Na_{1-x}Sn_2P_2$ was prepared by the solid-state reactions using $Na_3P$, Sn (Kojundo Chemical, 99.99%), and P (Kojundo Chemical, 99.9999%) as starting materials. To obtain $Na_3P$, Na (Sigma-Aldrichi, 99.9%) and P in a ratio of 3 : 1 were heated at 300 °C for 10 h in an evacuated quartz tube. A surface oxide layer of Na was mechanically cleaved before experiments. A stoichiometric mixture of $Na_3P$ : Sn : P = 1 : 3 : 2 was pressed into a pellet and heated at 400 °C for 20 h in an evacuated quartz tube. The obtained product was ground, mixed, pelletized, and heated again at 400 °C for 40 h in an evacuated quartz tube. The sample preparation procedures were conducted in an Ar-filled glovebox with a gas-purifier system or under vacuum. The obtained sample was stored in an Ar-filled glovebox because it is reactive in air and moist atmosphere.

The phase purity and the crystal structure of the samples were examined using synchrotron powder X-ray diffraction (SPXRD) performed at the BL02B2 beamline of the SPring-8 (proposal number of 2017B1283). The diffraction data was collected using a high-resolution one-dimensional semiconductor detector, multiple MYTHEN system (27). The wavelength of the radiation beam was determined to be 0.496916(1) Å using a $CeO_2$ standard. The crystal structure parameters were refined using the Rietveld method using the RIETAN-FP software (28). The crystal structure was visualized using the VESTA software (29).

Temperature ($T$) dependence of electrical resistivity ($\rho$) was measured using the four-terminal method with a physical property measurement system (PPMS; Quantum Design) equipped with a $^3$He-probe system. Magnetic susceptibility as a function of $T$ was measured using a superconducting quantum interference device (SQUID) magnetometer (Quantum Design MPMS-3) with an applied field of 10 Oe after both zero-field cooling (ZFC) and field cooling (FC). The specific heat ($C$) as a function of $T$ was measured using the relaxation method with PPMS.

Electronic structure calculations based on density functional theory were performed using the VASP code (30,31). The exchange-correlation potential was treated within the generalized gradient approximation using the Perdew−Becke−Ernzerhof method (32). The Brillouin zone was sampled using a 9 × 9 × 3 Monkhorst−Pack grid (33), and a cutoff of 350 eV was chosen for the plane-wave basis set. Spin-orbit coupling was included for the DFT calculation. Experimentally obtained structural parameters were employed for the calculation.

**Acknowledgement**

We thank R. Higashinaka and O. Miura for their experimental support. We thank K. Kuroki, H. Usui, T. Shibauchi, Y. Mizukami, and K. Ishihara for their fruitful discussion. This work was partly supported by Grants-in-Aid for Scientific Research (Nos. 15H05886, 15H05884, 16H04493, 17K19058, 16K17944, and 15H03693) and Iketani Science and Technology Foundation (No. 0301042-A), Japan.

**Author Contributions**

Y.G. performed sample preparation and characterization. Y.M. supervised the experimental work. Y.G., A.M., C.M., Y.K., and Y.M. conducted SXRD measurements. Y.G., T.D.M., Y.A. performed physical properties measurements. Y.G. performed DFT calculation. Y.G. and Y.M. wrote the manuscript with contributions from the other authors.

**Additional Information**

**Competing Interests:** The authors declare that they have no competing interests.

Table 1

Crystal structure parameters and reliability factors of $Na_{1-x}Sn_2P_2$ obtained from Rietveld refinement.[a]

| Lattice system | Trigonal | | | | | | |
|---|---|---|---|---|---|---|---|
| Space group | $R\bar{3}m$ (No. 166) | | | | | | |
| Lattice parameters | $a = 3.8847(2)$ Å | | | | | | |
| | $c = 27.1766(13)$ Å | | | | | | |
| | $\gamma = 120°$ | | | | | | |
| Atom | Site | Symmetry | $g^b$ | $x$ | $y$ | $z$ | $U$ (Å$^2$) |
| Na | 3a | −3m | 0.926(18) | 0 | 0 | 0 | 0.016(5) |
| Sn | 6c | 3m | 1 | 0 | 0 | 0.21181(6) | 0.0119(7) |
| P | 6c | 3m | 1 | 0 | 0 | 0.4063(2) | 0.0117(16) |
| $R_{wp}$ | 9.128% | | | | | | |
| $R_B$ | 3.648% | | | | | | |
| GOF | 3.6956 | | | | | | |

[a]Values in parentheses are standard deviations in the last digits.
[b]Site occupancies ($g$) of Sn and P sites were fixed at unity.

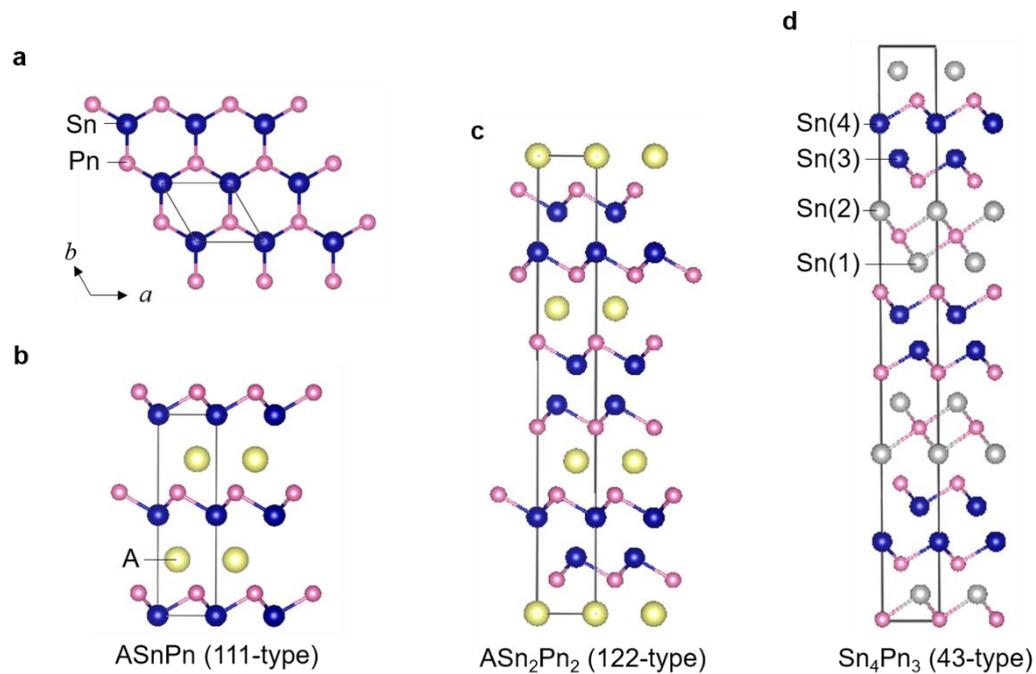

Figure 1

Schematic representations of crystal structure of SnPn-based layered compounds. (a) Honeycomb network of SnPn conducting layer. (b) Crystal structure of ASnPn (hexagonal $P6_3mc$ space group). (c) Crystal structure of ASn$_2$Pn$_2$ (trigonal $R\bar{3}m$ space group). (d) Crystal structure of Sn$_4$Pn$_3$ (trigonal $R3m$ space group). Here, A denotes the alkali metal or alkaline earth metal, and Pn denotes pnictogen. Black line represents the unit cell. For Sn$_4$Pn$_3$, there are two types of tin atom coordination in crystal structure. The Sn(1) and Sn(2) atoms are octahedrally coordinated by arsenic atoms only. The Sn(3) and Sn(4) atoms have a [3 + 3] coordination composed by three arsenic atoms from one side and three tin atoms beyond van der Waals (vdW) gap. To emphasize the similarity of SnPn layer with vdW gap, Sn(1) and Sn(2) atoms in Sn$_4$Pn$_3$ were drawn using different color from Sn(3) and Sn(4) atoms.

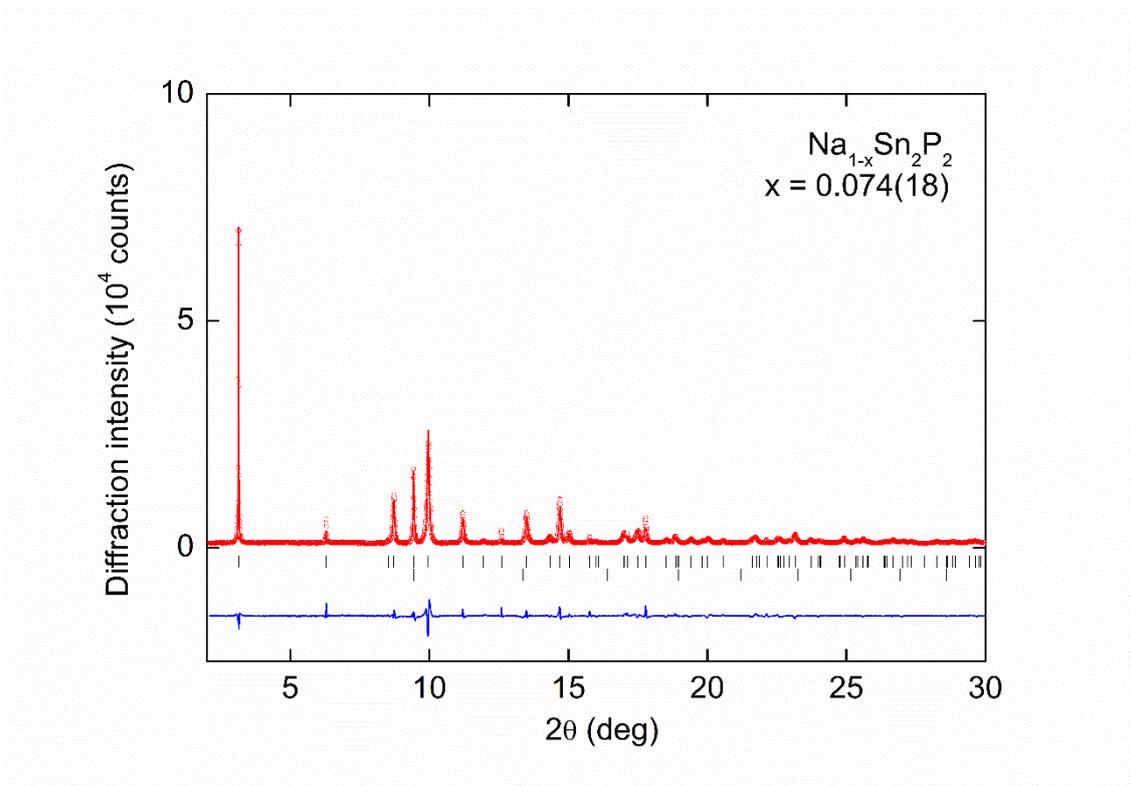

Figure 2
Synchrotron powder X-ray diffraction (SPXRD) pattern ($\lambda$ = 0.496916(1) Å) and the results of Rietveld refinement for $Na_{1-x}Sn_2P_2$. The circles and solid curve represent the observed and calculated patterns, respectively, and the difference between the two is shown at the bottom. The vertical marks indicate the calculated Bragg diffraction positions for $NaSn_2P_2$ (upper) and Na (lower).

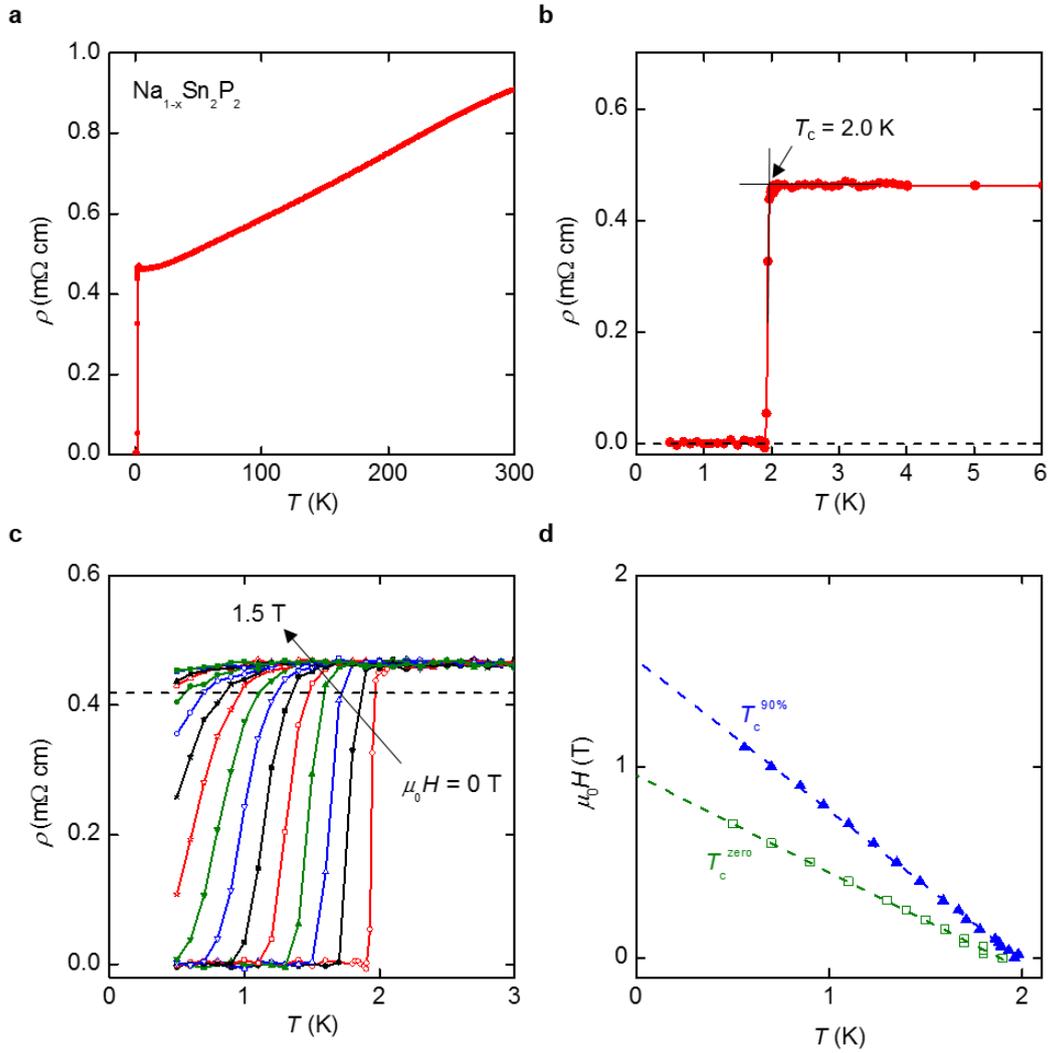

Figure 3

(a) Temperature ($T$) dependence of electrical resistivity ($\rho$) of Na$_{1-x}$Sn$_2$P$_2$. (b) $\rho$–$T$ data below 6 K. (c) $\rho$–$T$ data under magnetic fields up to 1.5 T with an increment of 0.1 T. Dashed line represents 90% of $\rho$ at 3 K. (d) Magnetic field–temperature phase diagram of NaSn$_2$P$_2$. Dashed lines represent the least-squares fits of data plots.

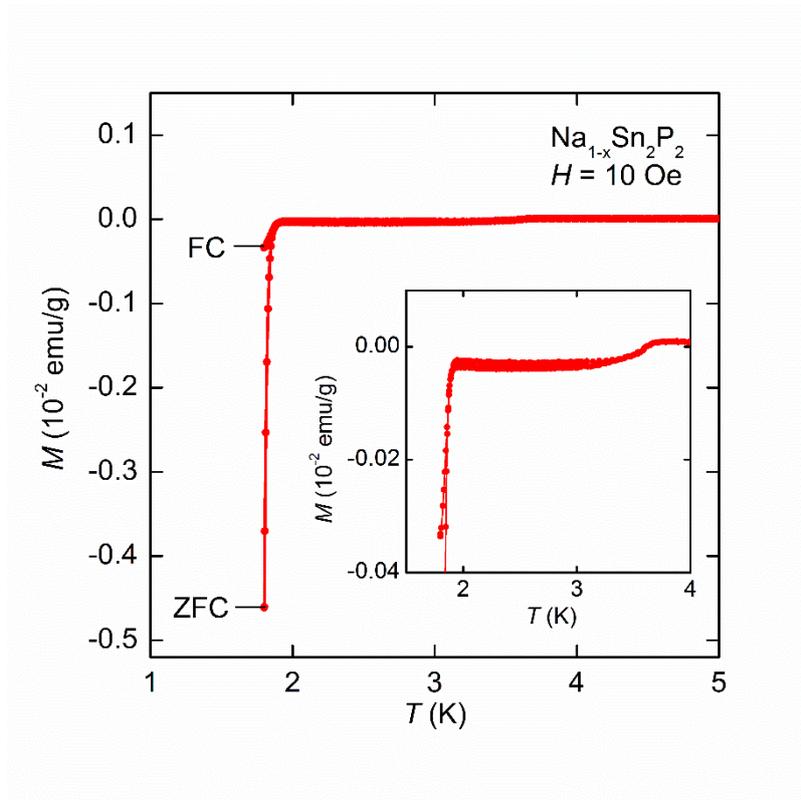

Figure 4

Magnetization ($M$) as a function of temperature ($T$) for Na$_{1-x}$Sn$_2$P$_2$ measured after both zero-field cooling (ZFC) and field cooling (FC). The inset shows enlarged view around superconducting transition.

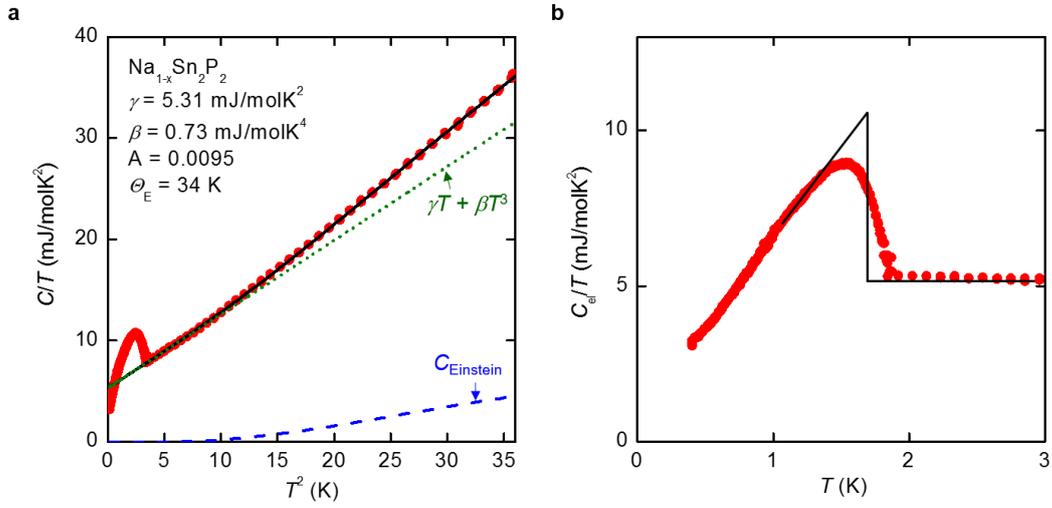

Figure 5

(a) Measured specific heat of Na$_{1-x}$Sn$_2$P$_2$ (red circles). Black line shows a fit to the experimental data above 2 K including electronic and phonon components (see text). Contributions from electrons and Debye phonon heat capacity ($\gamma T + \beta T^3$) and low-energy Einstein mode ($C_{\text{Einstein}}$) are denoted by green dotted line and blue dashed line, respectively. (b) Electronic specific heat of Na$_{1-x}$Sn$_2$P$_2$. Black solid line is used to estimate the specific heat jump ($\Delta C_{\text{el}}$) at $T_{\text{c}}$.

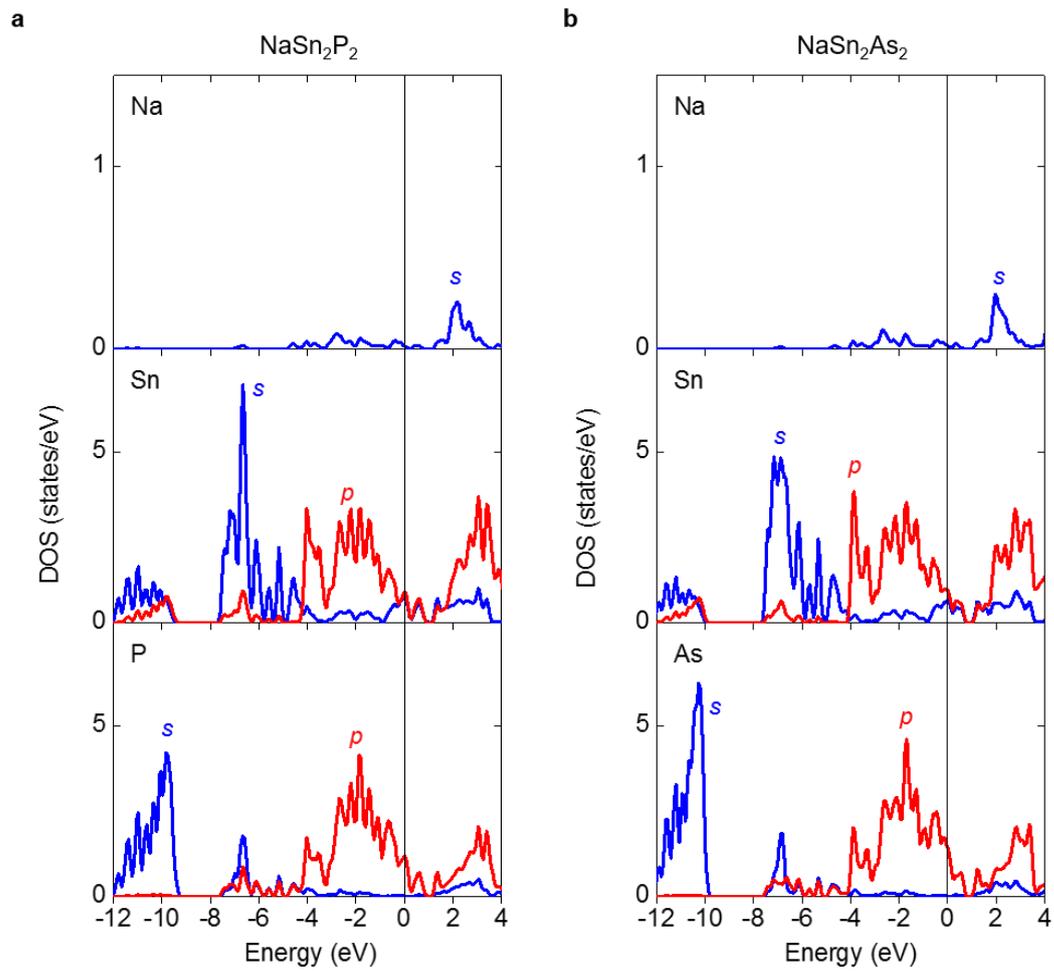

Figure 6

Partial density of states (DOS) of (a) NaSn$_2$P$_2$ and (b) NaSn$_2$As$_2$. The Fermi energy was set to 0 eV.